\documentclass[12pt]{WileyMSP-template}

\usepackage{geometry}
\usepackage{amssymb}
\usepackage{ragged2e}
\usepackage{xcolor} 
\usepackage[square,super,comma,sort&compress]{natbib}
\usepackage{multicol}
\usepackage{footnote}

\newenvironment{Figure}
{\par\medskip\noindent\minipage{\linewidth}}
{\endminipage\par\medskip}

\begin{document}


\title{\textbf{Proximity-tuned Magnetic and Transport \\Anomalies in All-epitaxial Fe$_{5-x}$GeTe$_2$/WSe$_2$ \\Van der Waals Heterostructures}}

\maketitle


\author{Hua Lv*},
\author{Tauqir Shinwari},
\author{Kacho Imtiyaz Ali Khan},
\author{Jens Herfort},
\author{Chen Chen},
\author{Joan M. Redwing},
\author{Mehak Loyal},
\author{Gerhard Jakob},
\author{Mathias Kläui},
\author{Achim Trampert},
\author{Bernat Mundet},
\author{Belén Ballesteros},
\author{Manfred Ramsteiner},
\author{Roman Engel-Herbert},
\author{Michael Hanke},
\author{João Marcelo J. Lopes*}

\begin{affiliations}

Dr. H. Lv, Dr. T. Shinwari, Dr. K. I. A. Khan, Dr. J. Herfort, Dr. A. Trampert, Dr. M. Ramsteiner, Prof. Dr. R. Engel-Herbert, PD Dr. M. Hanke, Dr. J. M. J. Lopes\\
Paul-Drude-Institut f\"ur Festk\"orperelektronik, Leibniz-Institut im Forschungsverbund Berlin e.~V., \\Hausvogteiplatz 5--7, 10117 Berlin, Germany\\
Email Address: hua.lv@pdi-berlin.de, lopes@pdi-berlin.de

Dr. C. Chen, Prof. Dr. J. M. Redwing \\
2D Crystal Consortium Materials Innovation Platform, Materials Research Institute, The Pennsylvania State University, University Park, PA 16802, United States

M.S. M. Loyal, Prof. Dr. G. Jakob, Prof. Dr. M. Kläui \\
Institute of Physics, Johannes Gutenberg University Mainz, Staudingerweg 7, 55128 Mainz, Germany

Dr. B. Mundet, Dr. B. Ballesteros \\
Catalan Institute of Nanoscience and Nanotechnology (ICN2), CSIC and The Barcelona Institute of Science and Technology (BIST), Campus UAB, Bellaterra, 08193 Barcelona, Spain

\end{affiliations}



\begin{abstract}
Van \justifying der Waals (vdW) heterostructures combining two-dimensional (2D) ferromagnets and semiconducting transition-metal dichalcogenides (TMDCs) offer highly promising opportunities for tailoring 2D magnetism through interfacial proximity effects, enabling unique physical phenomena inaccessible in 3D systems and achieving functionalities beyond conventional spintronics. However, current fabrication of vdW heterostructures still relies heavily on the manual stacking of exfoliated 2D flakes, leading to critical challenges in scalability, interfacial quality, thickness control and device integration. This work reports on the realization of all-epitaxial, high-quality Fe$_{5-x}$GeTe$_2$(FGT)/WSe$_2$ heterostructures exhibiting perpendicular magnetic anisotropy (PMA) and room-temperature ferromagnetism. The FGT/WSe$_2$ system demonstrates temperature-driven magnetic transitions, higher-order PMA contributions and large anisotropic magnetoresistance, highlighting sublattice-specific contributions to magnetic and transport properties. Notably, the FGT/WSe$_2$ heterostructures display unconventional physical phenomena, including thickness- and temperature-dependent sign reversal of exchange bias, a reversed thickness trend in the unconventional Hall effect, and a non-monotonic PMA-thickness dependence. These anomalies indicate pronounced interfacial contributions arising from proximity effects enhanced by epitaxial interface quality. Collectively, this study provides deep insights into the magnetic and transport properties of FGT/WSe$_2$ vdW heterostructures, establishing a scalable platform for exploring emergent 2D physics and advancing next-generation 2D spintronic technologies.
\end{abstract}



\justifying
 
\section{\label{sec:level1}Introduction}
The emergence of two-dimensional (2D) ferromagnets is driving a paradigm shift in magnetism research.\cite{Gong2017N,Huang2017N,Fei2018NM,Seo2020SA,Song2018S} While 3D magnetic materials have been extensively explored over the past century and underpin modern information technology, atomically thin 2D magnets are now at the forefront, promising to reveal novel magnetic phenomena and enable device concepts previously unattainable.\cite{Yang2022N,Huang2020NM,Gibertini2019NN,Tang2022NE,Zhang2025ACSnano,Lin2023ASCnano} Furthermore, magnetic heterostructures integrating different materials can provide great opportunities to engineer magnetic and transport properties, induce novel interfacial phenomena and achieve functionalities impossible in single materials.\cite{Huang2020NM,Gibertini2019NN,Jiang2019NE} However, conventional 3D heterostructures require strict lattice matching and often suffer from interlayer mixing and magnetic dead layers, which significantly degrade the interface quality and suppress interface-driven physics, representing a long-standing issue.\cite{Fert2024PRM,Kuepferling2023PRM,Manchon2019RMP,Hellman2017PRM} In contrast, van der Waals (vdW) heterostructures can be constructed from lattice-mismatched materials, which enables atomically sharp and strain-free interfaces,\cite{Geim2013N,Huang2020NM,Gibertini2019NN,Jiang2019NE,Lv2023Small} thereby significantly enhancing coupling between adjacent layers.\cite{Choi2022AS,Huang2020NM,Wu2022AM,Sierra2021NN,Yang2022N} 
In particular, 2D magnets and vdW heterostructures provide a promising platform for exploring phenomena that are absent or strongly suppressed in 3D systems, including: (i) proximity effects, offering an efficient but largely underexplored route to tune magnetic and transport properties; \cite{Huang2020NM,Choi2022AS} (ii) 2D exchange bias, an essential step for realizing 2D spintronic devices; \cite{Balan2024AM,Kumar2025hex,Cham2024AM,Xiao2024AM,Wu2022AM,Zhang2022SAhex} and (iii) unconventional spin transport and topological spin textures, which probe Berry curvature and hold strong promise for high-density racetrack memories.\cite{Schmitt2022CP,Ji2024AM,Liu2024AMsky,Lv2024SkrAFM,Casas2023AM,Zhang2022ASskyr} Investigating them not only deepens the fundamental understanding but also opens avenues for novel functionalities, effectively bridging conventional 3D spintronics with its 2D counterpart. However, systematic studies are still in their infancy, with many emerging 2D phenomena and underlying mechanisms yet to be explored.

Among various 2D magnets, Fe–Ge(Ga)–Te compounds stand out as prototypical material systems by combining high Curie temperature ($T_\mathrm{C}$) and intrinsic metallic behavior, both of which are essential for spintronic applications.\cite{Fei2018NM,Seo2020SA,May2019ACSnano,Lopes2021,Lv2023Small,Zhang2022NC,Shinwari2025} Within this family, metallic Fe$_5$GeTe$_2$ (FGT) exhibits a $T_\mathrm{C}$ near room temperature, positioning it as one of the leading material candidates.\cite{May2019ACSnano,Seo2020SA,Lv2023Small} Beyond high $T_\mathrm{C}$, perpendicular magnetic anisotropy (PMA) is critical for achieving low-power magnetization operation and high thermal stability, playing a key role in non-volatile magnetic memory technologies.\cite{Dieny2017RMP,Dieny2020NE} However, conventional 3D PMA materials, such as CoFeB/MgO and [Co/Pt]$_n$-like multilayers, require sub-nanometer thickness precision to maintain strong PMA, leading to significant challenges for scalable fabrication.\cite{Ikeda2010,Ikeda2012,Dieny2020NE} Identifying PMA materials without such thickness constraints is crucial for modern spintronics but remains an unmet goal.\cite{Dieny2020NE}
Previous studies of FGT in exfoliated flakes typically exhibit in-plane magnetic anisotropy (IMA),\cite{Miao2023ACSnano,Deng2022,Zhang2020PRB} thereby strongly limiting their spintronic applications. Furthermore, to build FGT-based heterostructures, 2D semiconducting transition-metal dichalcogenides (TMDCs) are particularly attractive partners due to their strong spin–orbit coupling (SOC) and unique optical response.\cite{Sierra2021NN} Fundamentally, FGT/TMDC heterostructures could offer a platform to exploit proximity-induced phenomena without introducing additional magnetic and transport signals on FGT, including PMA modulation and skyrmion formation.\cite{Kim2021ACSnano} From an application perspective, such heterostructures promise large spin-orbit torque (SOT) effects \cite{Shin2022cis} and optical spin injection, providing an ideal platform for combining spin-orbitronic and opto-spintronic functionalities.\cite{Sierra2021NN} 
However, most demonstrations of 2D magnets and related vdW heterostructures have primarily relied on exfoliated flakes assembled manually, which are typically limited to micrometer dimensions and thus incompatible with scalable device fabrication.\cite{Zhang2025ACSnano,Lin2023ASCnano,Dong2024AM} For both fundamental studies and practical applications, achieving large-scale epitaxial growth of 2D magnets and vdW heterostructures is therefore essential.\cite{Yang2022N,Zhang2025ACSnano,Lin2023ASCnano} Indeed, following the investigation of various 2D materials in bulk crystal form, achieving large-scale epitaxial synthesis is highly demanded for future applications.\cite{Zhu2023NatNano,Lin2023ASCnano,Li2024NM,Liu2024NN,Lv2023Small} However, despite recent progress in scalable fabrication of vdW heterostructures,\cite{Lv2023Small,Shinwari2025,Guillet2023} all-epitaxial growth of 2D magnet/TMDC heterostructures with near-room-temperature magnetism have not yet been realized, representing a critical bottleneck for advancing 2D spintronic applications.\cite{Sierra2021NN}

In this work, we report the realization of large-area, all-epitaxial FGT/TMDC vdW heterostructures, here using WSe$_2$ as the TMDC, which exhibit atomically sharp interfaces and room-temperature ferromagnetism with PMA. The observed temperature-driven magnetic transitions, higher-order PMA and pronounced anisotropic magnetoresistance (AMR) highlight sublattice-specific contributions to the magnetic and transport properties. Strikingly, the FGT/WSe$_2$ heterostructures reveal a series of unconventional phenomena associated with their 2D nature, including a temperature- and thickness-dependent sign reversal of exchange bias, a non-monotonic PMA–thickness dependence, and a reversed thickness trend in the unconventional Hall effect (UHE). These phenomena indicate pronounced interfacial contributions, suggesting a strong proximity effect enabled by the high FGT/WSe$_2$ interface quality. This work not only provides deeper insights into the magnetic and transport properties of FGT/WSe$_2$ vdW heterostructures but also establishes an excellent platform for exploring emergent 2D physics and developing next-generation spintronic technologies.

\section{Results and Discussion}
\subsection{\label{sec:leve21} Structural properties}
Molecular beam epitaxy (MBE) was employed to grow FGT films with thicknesses ($t_\mathrm{FGT}$) of 6, 12 and 17 nm directly on single-crystalline, continuous monolayer–bilayer thick WSe$_2$ synthesized by chemical vapor deposition (CVD) on Al$_2$O$_3$(0001).\cite{Zhu2023NatNano} The FGT growth process was continuously monitored via in-situ reflection-high-energy-electron-diffraction (RHEED, Figure S1, Supporting Information), which confirmed the epitaxial growth of FGT films having a smooth 2D surface. The FGT films were capped \textit{in-situ} with a Te layer (5 nm) deposited at room temperature. Fe$_{5-x}$GeTe$_2$ is usually a non-stoichiometric system,\cite{May2019PRM,Ershadrad2022JPCL} with x $\approx$ 0.2 in our case.\cite{Lv2023Small}

\begin{Figure}
\vspace*{0.5cm}
\centering
\includegraphics*[width=0.95\textwidth]{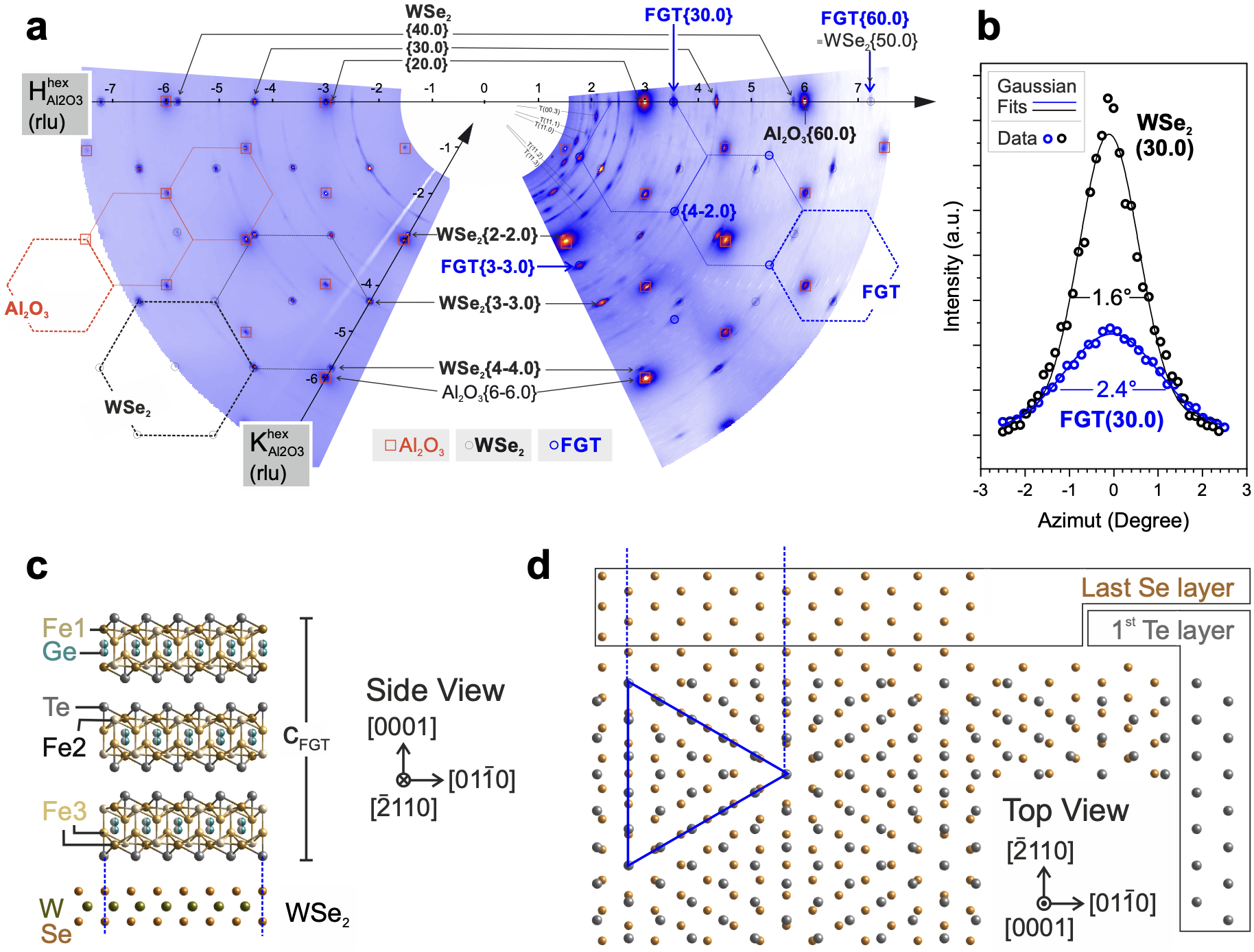}
\captionof{figure}{a) In-plane reciprocal space maps: (left) bare WSe$_2$/Al$_2$O$_3$(0001) and (right) 17 nm FGT/WSe$_2$/ Al$_2$O$_3$(0001). Due to trigonal symmetry, the H and K axes form an angle of 60$^{\circ}$. b) Angular scans intersecting the WSe$_2$ and FGT(30.0) reflections in the space map shown in the right panel of a). c,d) Schematics of the structural configuration of the synthesized FGT/WSe$_2$ stack according to the epitaxial alignment determined by GID: c) side-view, d) top-view. For simplicity, d) shows only the last, topmost Se layer in WSe$_2$ and the first Te layer in FGT. The blue triangle indicates the coincidence-site lattice matching between WSe$_2$ and FGT.}
\label{Fig:GID}
\vspace*{0.5cm}
\end{Figure}

The structural properties of the samples were investigated in detail by synchrotron-based grazing incidence diffraction (GID), a tool of choice for probing in-plane properties such as strain and crystallographic orientation of 2D layers.\cite{Schumann2014} Due to the evanescent X-ray wave field in the crystal, GID becomes an extremely surface-sensitive approach. \textbf{Figure \ref{Fig:GID}a} shows two reciprocal space maps probed by GID. They cover a relatively large area, i.e., azimuthally a range of $\pm$ 35$^{\circ}$ and in radial direction scattering vectors far beyond the in-plane Al$_2$O$_3$ (60.0) reflection. The reciprocal space map shown left was obtained for a pristine WSe$_2$/Al$_2$O$_3$(0001) samples, while the right one was taken for a 17 nm FGT/WSe$_2$/Al$_2$O$_3$(0001) sample. The former serves as a reference for the sample with an additional FGT layer on top (right). To disentangle the different scattering contributions from sapphire, WSe$_2$ and FGT, we have marked the respective positions and connected some of them with dashed lines. This gives a clear view of the crystallographic orientation, with the [10.0] substrate direction coinciding exactly with that of FGT and WSe$_2$. Despite the weak vdW bonding, a well-defined in-plane epitaxial relation between the two layers is preserved.
This epitaxial alignment is also evidenced by azimuthal scans, as depicted in Figure \ref{Fig:GID}b for the FGT(30.0) and WSe$_2$(30.0) in-plane reflections. Both angular widths, i.e. the full width at half maximum values (1.6$^{\circ}$ for the WSe$_2$ layer and 2.4$^{\circ}$ for FGT) are rather small considering what is usually measured for epitaxially grown vdW materials.\cite{Lv2023Small,Nakano2017} Note that, since the FGT/WSe$_2$ stack was covered by a Te capping layer, we could also probe the signature of metallic tellurium by a set of reflections rather close to the center.

Taking the radial distance for a set of reflections for FGT and WSe$_2$, we could precisely determine the in-plane lattice parameters: for FGT, a value \textit{a}\textsubscript{FGT} = 4.032(10) \AA{} was found, while for WSe$_2$ the values \textit{a}\textsubscript{WSe$_2$} = 3.299(10) \AA{} and \textit{a}\textsubscript{WSe$_2$} = 3.273(10) Å were obtained for pristine and FGT-overgrown films, respectively. These values are in good agreement with what has been reported for bulk crystals and thin films grown on different substrates.\cite{Lv2023Small,Nakano2017,May2019ACSnano,Ribeiro2022,Mortelmans2019,Voss1999} The smaller in-plane lattice constant observed in WSe$_2$ overgrown by FGT may be related to defect formation in WSe$_2$ during the MBE process. In particular, heating the sample in the ultra-high vacuum (UHV) environment of MBE may cause some Se to desorb, resulting in vacancy formation and local lattice contractions.\cite{Yang2019NRL} This, in turn, may lead to a small reduction in the average lattice constant, which GID can detect.\cite{Schumann2013} Nevertheless, the clear observation of several GID reflections associated with WSe$_2$ serves as concrete evidence that the TMDC layer preserves its overall structure after FGT growth. This is further supported by Raman spectroscopy measurements (Figure S2, Supporting Information). Finally, it is important to point out that GID reveals a coincidence-site lattice matching between FGT and WSe$_2$, with a resulting ratio 5$\textit{a}\textsubscript{FGT}$ $\approx$ 6$\textit{a}\textsubscript{WSe$_2$}$. This likely contributes to the strong in-plane registry between both materials, as previously reported for other epitaxial vdW systems.\cite{Boschker2015} The epitaxial orientation and coincidence-site lattice matching between the FGT and WSe$_2$ layers are depicted schematically in Figure \ref{Fig:GID}c,d. Furthermore, GID reveals very similar structural properties for FGT films of different thicknesses (Figure S3, Supporting Information).

\begin{figure*}[t!]
\centering
\includegraphics*[width=0.8\textwidth]{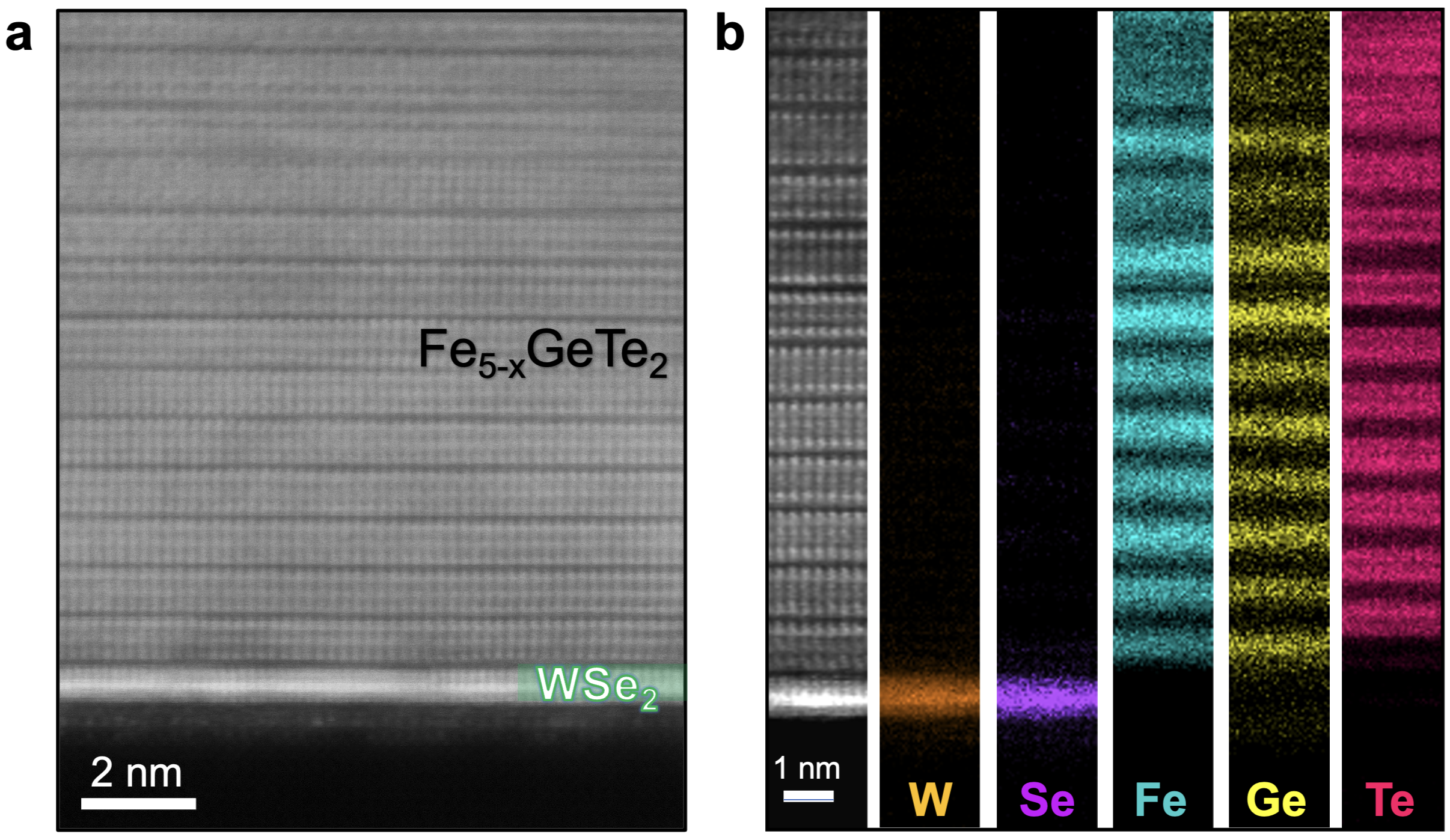}
\captionof{figure}{a) STEM cross-section image of a 12 nm thick FGT film grown on WSe$_2$/Al$_2$O$_3$(0001). b) STEM image (left) and the EELS compositional maps for W (orange), S (purple), Fe (blue), Ge (yellow) and Te (red) obtained from the same region.} 
\label{Fig:STEM}
\end{figure*}

Scanning transmission electron microscopy (STEM) was also employed to investigate the structural properties of the layers. \textbf{Figure \ref{Fig:STEM}a} shows a high-angle annular dark-field (HAADF)-STEM cross-sectional image of the FGT/WSe$_2$ heterostructure acquired in a region where the WSe$_2$ film is monolayer thick. A similar image obtained in a region of bilayer WSe$_2$ is depicted in Figure S4, Supporting Information. The layered structure of the FGT film is clearly visible, in that single FGT layers are separated by vdW gaps. This is consistent with previous TEM characterization of FGT bulk crystals and thin films. \cite{Lv2023Small,May2019ACSnano,Ribeiro2022} By analyzing images obtained from different regions, it was found that the FGT film comprises not only Fe$_5$GeTe$_2$ layers (with a monolayer thickness of ~1 nm), but also with thicknesses ranging from 1 to 0.8 nm. The latter value, for example, is in very good agreement with the thickness of a Fe$_3$GeTe$_2$ single layer.\cite{May2016PRB} The average thickness of the FGT single layers obtained from TEM was found to be 0.93 nm. We attribute the stacking disorder in the FGT film — namely, coexistence of layers with slightly different thicknesses and potentially different Fe compositions — to the relatively low substrate temperature employed for the growth of FGT (260°C). Previous studies on MBE growth of FGT films on substrates other than WSe$_2$ (e.g., Al$_2$O$_3$, Bi$_2$Te$_3$) have shown that higher growth temperatures and/or post-growth annealing favor stabilization of Fe$_{5-x}$GeTe$_2$ phases with a higher Fe concentration, such as Fe$_5$GeTe$_2$ (i.e., $x$ = 0).\cite{Ribeiro2022,Zhou2023PRM} In this study, a low growth temperature was intentionally chosen to mitigate the possible intermixing at the FGT/WSe$_2$ interface. This strategy proved successful, as the electron energy loss spectroscopy (EELS) mappings show a fairly homogeneous distribution of each element with no intermixing between adjacent layers (see Figure \ref{Fig:STEM}b). As expected for FGT (see Figure \ref{Fig:GID}c), Fe and Ge are usually positioned at the center of the FGT layers between the outer Te slabs, while W and Se are present where WSe$_2$ is located. Note that a region with different contrast can be observed directly below the WSe$_2$ film at the WSe$_2$/sapphire interface (see Figures \ref{Fig:STEM}a and S4). This is due to the formation of a Se passivation layer, as discussed in detail previously.\cite{Zhu2023NatNano}

\subsection{\label{sec:leve22} Magnetic properties}
\begin{figure*}
\centering
\includegraphics*[width=1\textwidth]{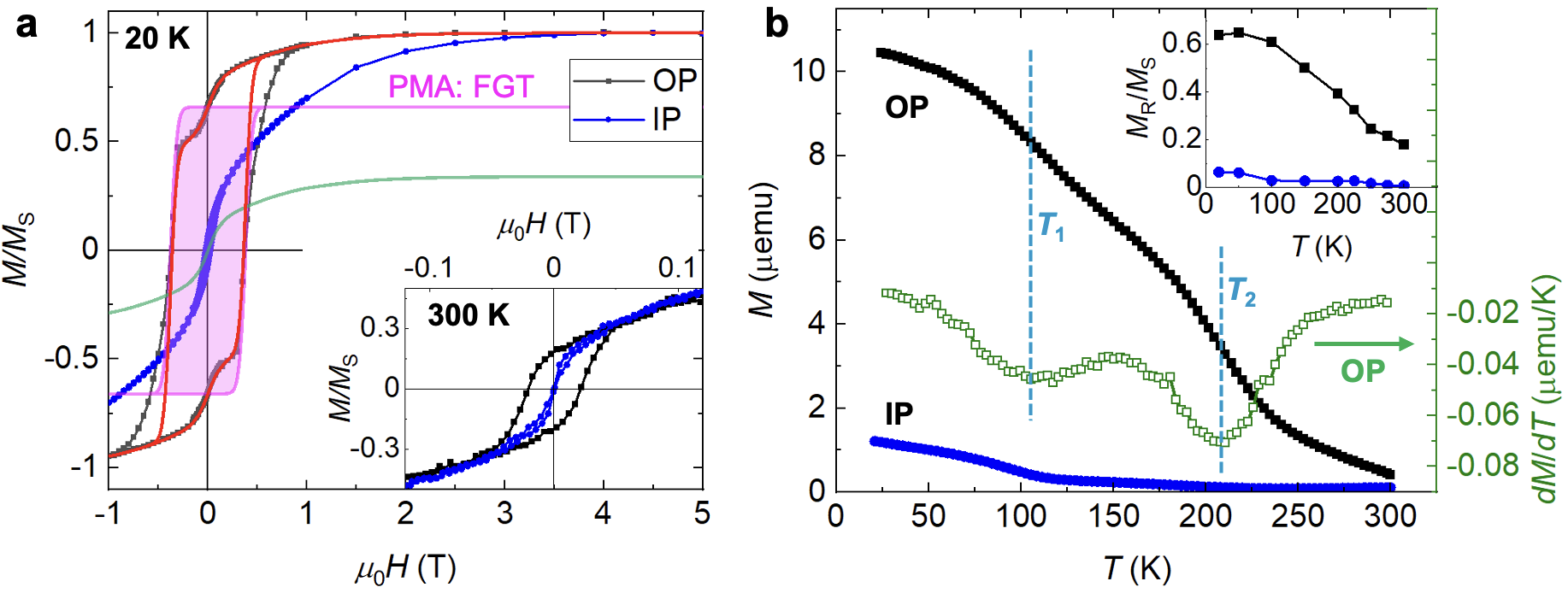}
\captionof{figure}{Magnetic characterization of a 17 nm FGT/WSe$_2$ vdW heterostructure using SQUID magnetometry.
a) Normalized magnetization $M/M_\mathrm{S}$ as a function of out-of-plane (OP, black) and in-plane (IP, blue) magnetic fields at 20 K. Red curves are fits based on Equation (\ref{eq:hys}), comprising a square-shaped PMA component (pink) and a zero-coercivity contribution (green). Inset: normalized hysteresis loops at 300 K.
b) Temperature-dependent magnetization $M$ during zero-field warming. The green curve shows the first derivative $dM/dT$ in the OP configuration. Inset: $M_\mathrm{R}/M_\mathrm{S}$ extracted from the hysteresis loops at each temperature.} 
\label{fig:SQUID}
\end{figure*}

The magnetic properties of FGT/WSe$_2$ vdW heterostructures were investigated using superconducting quantum interference device (SQUID) magnetometry. \textbf{Figure \ref{fig:SQUID}a} presents the hysteresis loops normalized by the saturation magnetization ($M_\mathrm{S}$) after subtracting a linear background (see raw data in Figure S5, Supporting Information), for a 17 nm thick FGT sample measured at 20 K with a magnetic field ($H$) applied along the out-of-plane (OP, black) and in-plane (IP, blue) directions. The markedly higher coercive field ($H_\mathrm{C}$) and lower saturation field ($H_\mathrm{S}$) observed in the OP compared to the IP configuration indicate a strong PMA. Notably, the OP hysteresis loop exhibits an unconventional shape with a kink near the zero field (see also Figure S6, Supporting Information). To interpret this feature, an empirical model was developed based on the Boltzmann sigmoid function (Text S1, Supporting Information):
\begin{equation}
\label{eq:hys}
\frac{M(H)}{M_\mathrm{S}} = 1 - \frac{2}{1 + e^{\frac{H \pm H_\mathrm{c}}{(H_\mathrm{s} \pm H_\mathrm{c})/4}}}
\end{equation}
Equation (\ref{eq:hys}) provides a concise description of the hysteresis loop using only the directly measurable parameters of $H_\mathrm{C}$ and $H_\mathrm{S}$. This approach offers significant advantages over more complex models, such as the Jiles-Atherton model based on the Langevin function,\cite{Jiles1984} particularly for quantitative analyses and simulations. The OP hysteresis loop can be well fitted by Equation (\ref{eq:hys}) (red curves) when incorporating a square-shaped phase (pink curves) and a zero-coercivity contribution (green curve) (see Text S1, Supporting Information). The square-shaped phase is attributed to the strong PMA of FGT and closely matches the anomalous Hall effect (AHE) signal (see the next section). The zero-coercivity contribution, given its absence in the AHE measurement, is unlikely to originate from the metallic FGT phase. It may arise from the formation of oxidized FGT (O-FGT), a possibility that will be further discussed alongside the transport measurements.

Figure \ref{fig:SQUID}b illustrates the temperature-dependent magnetization measured during zero-field warming up after setting the initial magnetization with a 1 T field. The inset displays the normalized remanent magnetization ($M_\mathrm{R}$/$M_\mathrm{S}$), with $M_\mathrm{R}$ obtained at zero field from the $M-H$ curves (Figure \ref{fig:SQUID}a). The larger $M_\mathrm{R}$ in the OP configuration indicates the persistence of PMA and ferromagnetic ordering up to room temperature. Both the OP $M-T$ curve (black) and its derivative $dM/dT$ (green) exhibit two clear magnetic transitions at temperatures $T_1 \approx$ 110 K and $T_2 \approx$ 220 K. The transition at $T_1$ is associated with the Fe(1) sublattice (the Fe atom near the vdW gap, see Figure \ref{Fig:GID}c), which remains magnetically ordered below 110 K but undergoes strong fluctuations in the intermediate temperature range.\cite{May2019ACSnano,May2019PRM} Previously, a magnetostructural phase transition from a metastable phase to a stable phase has been reported in FGT.\cite{Schmitt2022CP} However, this transition only occurs during the initial cooling process and remains permanent.\cite{Schmitt2022CP} Consequently, the magnetic transition at $T_1$ in our sample cannot be attributed to this phenomenon, as the measurement was performed starting from a low temperature of 4.3 K. The magnetic transition at $T_2$ is likely associated with the change in magnetic ordering of Fe(2).\cite{May2019ACSnano,May2019PRM} A minor contribution may also arise from the presence of the Fe$_3$GeTe$_2$ phase. However, given its small fraction, the magnetism should be predominantly governed by the FGT phase, as further supported by transport measurements (see next section). The ferromagnetism above $T_2$ is primarily attributed to the long-range ferromagnetic order of Fe(3).\cite{May2019ACSnano,May2019PRM} This gradual change in the magnetic ordering of different Fe sublattices may also explain the slow decay of FGT magnetization at high temperatures (e.g., above 250 K in Figure \ref{fig:SQUID}b). In contrast, the 2D ferromagnets Fe$_3$GeTe$_2$ \cite{Fei2018NM,Lopes2021} and Fe$_3$GaTe$_2$ \cite{Shinwari2025} do not exhibit anomalous magnetic transitions in their $M-T$ and $dM/dT$ curves, which is consistent with the absence of reported evidence for distinct sublattice-specific ordering transitions. This behavior aligns with the sharp decrease in magnetization observed in Fe$_3$GeTe$_2$ \cite{Fei2018NM,Lopes2021} and Fe$_3$GaTe$_2$ \cite{Shinwari2025} near their $T_\mathrm{C}$. Consequently, these anomalies in temperature-dependent magnetic properties constitute a unique characteristic of FGT, with their occurrence indicating the high-quality formation of the expected FGT phase. This conclusion is further substantiated by the observation that the $T_1$ reported in various FGT samples remains consistent irrespective of synthesis methods of epitaxy or exfoliation.\cite{Miao2023ACSnano,Deng2022,Zhang2020PRB}
The successful realization of epitaxial FGT films on a TMDC template that exhibit PMA without compromising $T_\mathrm{C}$ offers a highly promising platform for the advancement of 2D spintronics.

\subsection{\label{sec:leve23} Anomalous Hall effect (AHE)}
To further investigate the magnetic and transport behaviors, AHE measurements were performed due to their high sensitivity to magnetization in ultrathin films.\cite{Nagaosa2010} Large-area Hall-bar devices, benefiting from our scalable synthesis, were used (Figure S7, Supporting Information). \textbf{Figure \ref{fig:AHE}}a,b displays the AHE curves of a 6 nm thick FGT/WSe$_2$ heterostructure, with low- and high-temperature data presented separately due to different AHE amplitudes.
At low temperatures (Figure \ref{fig:AHE}a), the square-shaped AHE loops originate from the FGT layer and confirm the presence of strong PMA. The absence of kink-like features in the AHE data indicates that the zero-coercivity component in the SQUID signal (green curve, Figure \ref{fig:SQUID}a) likely arises from the insulating surface oxide, previously reported to be antiferromagnetic.\cite{Liang2023PRM} As oxidation is difficult to avoid entirely, even with a protective capping layer, this result highlights the complementary roles of AHE and SQUID measurements: the former selectively probes the conductive FGT layer, while the latter captures the magnetic response of the entire sample. At high temperature (Figure \ref{fig:AHE}b), the AHE loops become rounded with reduced coercivity, indicating a transition from hard to soft ferromagnetism. The AHE signal vanishes near 340 K, above which only a linear ordinary Hall effect (OHE) remains.

\begin{Figure}
\vspace*{0.5cm}
\centering
\includegraphics[width=0.99\textwidth]{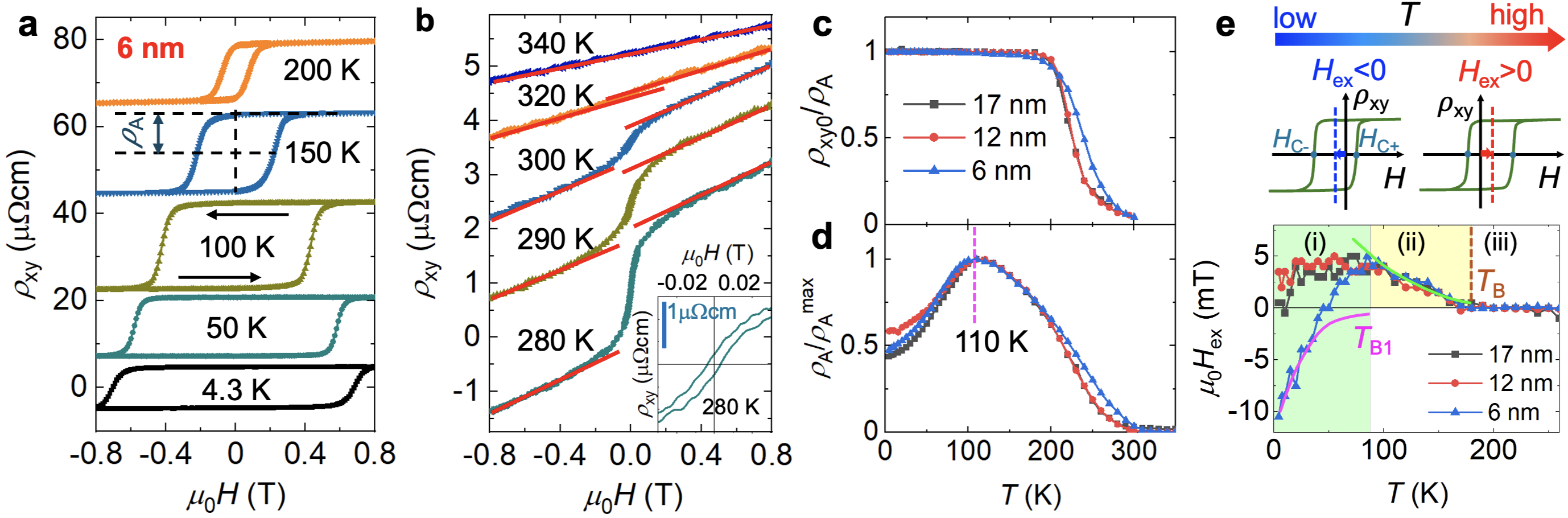}
\captionof{figure}{AHE and exchange bias effect in FGT/WSe$_2$ vdW heterostructures. a,b) Hall resistivity ($\rho_{xy}$) measured during upward and downward sweeps of the OP magnetic field at different temperatures, with sweep directions indicated by black arrows. The red lines in (b) are guides to the eye for the linear dependence in the saturation region. c,d) Temperature-dependent $\rho_{xy0}$/$\rho_\mathrm{A}$ and $\rho_\mathrm{A}$/$\rho_\mathrm{A}^{max}$ for samples with different FGT thicknesses, where $\rho_{xy0}$ is the zero-field $\rho_{xy}$ and $\rho_\mathrm{A}^{max}$ is the maximum $\rho_\mathrm{A}$. e) Top: Schematic of the sign reversal of the exchange bias controlled by temperature. Bottom: Temperature-dependent $H_{ex}$, with pink and green lines guiding the interfacial and bulk exchange coupling contributions, respectively.} 
\label{fig:AHE}
\vspace*{0.5cm}
\end{Figure}

Figure \ref{fig:AHE}c displays the normalized remanent AHE resistivity, $\rho_{xy0}$/$\rho_\mathrm{A}$, as a function of temperature for samples with different FGT thicknesses, where $\rho_{xy0}$ and $\rho_\mathrm{A}$ are the AHE resistivities at zero and saturation fields (at 0.8 T with the OHE contribution subtracted), respectively. The $\rho_{xy0}$/$\rho_\mathrm{A}$ remains constant up to approximately 200 K and then gradually decreases, however sustaining magnetic order up to 300 K. Figure \ref{fig:AHE}d shows the temperature-dependent $\rho_\mathrm{A}$/$\rho_\mathrm{A}^{max}$, where $\rho_\mathrm{A}^{max}$ represents the maximum value of $\rho_\mathrm{A}$ obtained at $T_1$ = 110 K. Below 110 K, the increase in $\rho_\mathrm{A}$ with increasing temperature can be mainly attributed to transport mechanisms, including intrinsic contribution of Berry curvature and extrinsic phonon-enhanced spin-dependent scattering.\cite{May2019PRM,Nagaosa2010} At 110 K, the peak in $\rho_\mathrm{A}$ is attributed to the magnetic unordering of Fe(1), while the subsequent decline at higher temperatures reflects the progressive reduction in overall magnetization.\cite{May2019PRM,Schmitt2022CP} Combining SQUID (Figure \ref{fig:SQUID}) and AHE (Figure \ref{fig:AHE}a-d) results, we estimate that the $T_\mathrm{C}$ of our FGT/WSe$_2$ heterostructures lies slightly above room temperature, with negligible dependence on FGT thickness. Notably, the $T_\mathrm{C}$ obtained in our epitaxial samples is comparable to the values reported for exfoliated FGT flakes ($\approx$ 270 - 310 K).\cite{Schmitt2022CP,Deng2022} Furthermore, the $T_1$ temperature (110 K) determined from $\rho_\mathrm{A}^{max}$ agrees well with that of bulk materials \cite{May2019PRM} and, in our case, is identical for all samples (Figure \ref{fig:AHE}d). This indicates that the Fe(1) unordering temperature is robust and not strongly dependent on film thickness.

\subsection{\label{sec:leve24} Sign reversal of exchange bias effect}
The exchange bias effect involving 2D ferromagnets has emerged as a compelling research topic due to its fundamental significance and promising spintronic applications.\cite{Balan2024AM,Cham2024AM,Xiao2024AM,Wu2022AM,Zhang2022SAhex} Unexpectedly, an exchange bias effect is observed in our FGT/WSe$_2$ heterostructures at low temperatures after cooling without additional field (see Methods), manifesting as a clear shift of the AHE loops' center from zero field, as schematically illustrated in the top panel of Figure \ref{fig:AHE}e. The bottom panel of Figure \ref{fig:AHE}e displays the extracted exchange bias field ($H_\mathrm{ex}$) as a function of temperature for FGT films of different thicknesses, where $H_\mathrm{ex}$ = ($H_\mathrm{C+}$+$H_\mathrm{C-}$)/2 with $H_\mathrm{C\pm}$ denoting the positive and negative coercive fields, respectively (see Figures S8 and S9, Supporting Information). Remarkably, $H_\mathrm{ex}$ exhibits a sign reversal between negative and positive depending on both temperature and FGT thickness, a highly unusual behavior rarely reported in conventional 3D magnetic systems. Three distinct regimes can be identified: (i) a thickness-dependent $H_\mathrm{ex}$ below 80 K with a positive temperature coefficient, (ii) a thickness-independent positive $H_\mathrm{ex}$ between 80 and 160 K with a negative temperature coefficient, and (iii) a vanishing $H_\mathrm{ex}$ above the Blocking temperature ($T_\mathrm{B}$) of 160 K.
The thickness-dependent $H_\mathrm{ex}$ dominant in region (i) likely originates from an interfacial exchange bias, consistent with Meiklejohn–Bean (MB) model with a negative $H_\mathrm{ex} \propto$ 1/$t_\mathrm{FGT}$.\cite{Nogues1999,kiwi2001,OGrady2010JMMM,Lv2019} This interfacial exchange bias may arise from the exchange coupling at the O-FGT/FGT interface, where an antiferromagnetic order and an induced exchange bias effect have been reported in naturally oxidized Fe$_5$GeTe$_2$,\cite{Liang2023PRM} Fe$_3$GeTe$_2$ \cite{Wu2022AM,Balan2024ACSnano} and Fe$_3$GaTe$_2$.\cite{Shao2024} 
Although our films do not show a distinct interface that would clearly distinguish oxidized and non-oxidized FGT regions in STEM characterizations, the possibility of non-uniform oxidation across the layers cannot be excluded. In contrast, the positive $H_\mathrm{ex}$ in region (ii) cannot be fully explained by the conventional exchange bias theory.\cite{Nogues1999,kiwi2001,OGrady2010JMMM,Lv2019} Instead, considering the thickness independence of both $H_\mathrm{ex}$ and $T_\mathrm{B}$ in this temperature region, this behavior likely arises from the intrinsic bulk properties of FGT. Indeed, a positive $H_\mathrm{ex}$ has recently been observed in Fe$_3$GeTe$_2$ flakes and was attributed to the coexistence of stable and frustrated magnetism at the sublayer surface.\cite{Hu2022} This suggests that the positive $H_\mathrm{ex}$ may be intrinsically associated with the 2D nature of FGT.\cite{Hu2022,Phan2023JAC} Consequently, the unusual exchange bias can be attributed to the interplay between two contributions: a negative $H_\mathrm{ex}$ occurring at the O-FGT/FGT interface with an estimated $T_\mathrm{B1} \approx$ 100 K (pink line in Figure \ref{fig:AHE}e), and a positive $H_\mathrm{ex}$ emerging from the interaction between FGT sublayers with a $T_\mathrm{B}$ = 160 K (green line in Figure \ref{fig:AHE}e). However, a comprehensive understanding of this unconventional exchange bias remains elusive.\cite{Balan2024AM,Hu2022,Phan2023JAC} The complex structural and magnetic properties of O-FGT/FGT are likely sensitive to factors such as the oxidation degree, the chemical composition and the local electronic structure,\cite{Guo2022,Phan2023JAC,Balan2024ACSnano} which require a systematic study beyond the scope of this paper. Nevertheless, it can be anticipated that the exchange bias in our FGT samples can be further tuned by precise control of O-FGT formation (e.g., controlling air exposure time, intently introducing O$_2$ and growing adjacent oxide layer \cite{Liang2023PRM,Balan2024ACSnano}). Finally, this intrinsic, single-material-induced self-exchange bias with a tunable sign provides a unique platform for exploring interfacial magnetism and developing 2D spintronic devices.

\subsection{\label{sec:leve25} In-plane Magnetotransport}
\begin{figure*}[t!]
\includegraphics*[width=0.99\textwidth]{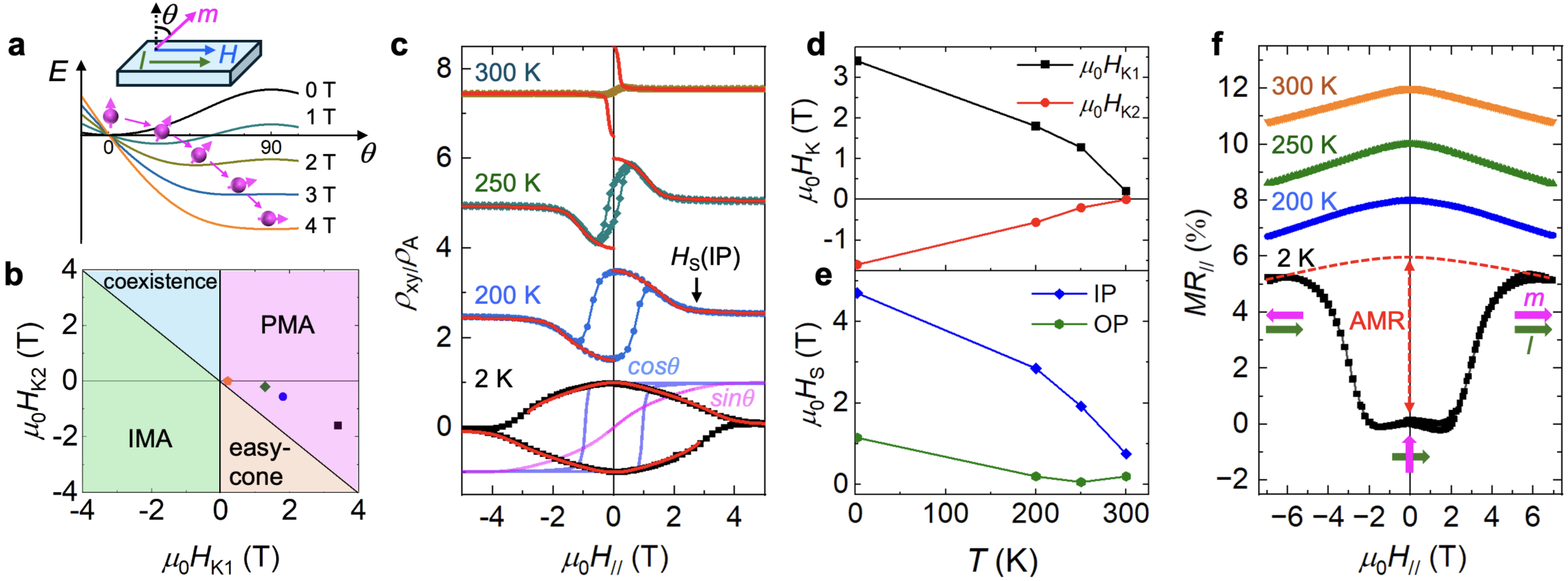}
\captionof{figure}{In-plane magnetotransport properties of a 6-nm-thick FGT/WSe$_2$ vdW heterostructure.
a) Schematic of magnetization reorientation (pink arrows with spheres) in a PMA material under an applied in-plane magnetic field $H_{\parallel}$, determined by the minima of the free-energy profile (colored curves). Inset: measurement configuration.
b) Phase diagram of possible magnetic states as functions of $H_\mathrm{K1}$ and $H_\mathrm{K2}$, symbols mark values extracted from the fits in c).
c) $\rho_{xy}$/$\rho_\mathrm{A}$ versus $H_{\parallel}$, with $\rho_\mathrm{A}$ obtained from AHE measurements (Figure \ref{fig:AHE}). The red curves are fits based on Equation (\ref{eq:PMAenergy}).
d) Temperature dependence of $H_\mathrm{K1}$ and $H_\mathrm{K2}$ extracted from fits in c).
e) Temperature dependence of $H_\mathrm{S}$ obtained from AHE loops with OP (Figure \ref{fig:AHE}a,b) and IP (see estimation in c) field orientations.
f) In-plane longitudinal magnetoresistance $MR_{\parallel}$ versus $H_{\parallel}$ at different temperatures; the red line indicates the method used to estimate the AMR value.}
\label{fig:IPtransport}
\end{figure*}

To gain a deeper insight into the magnetic anisotropy in FGT, magnetotransport measurements were also performed by applying an in-plane magnetic field ($H_{\parallel}$), as illustrated in \textbf{Figure \ref{fig:IPtransport}}a.\cite{Tang2022NE} In this configuration, the magnetization vector (pink arrows in Figure \ref{fig:IPtransport}a) continuously moves from the OP to IP direction as $H_{\parallel}$ increases, settling at an angle ($\theta$) determined by the free energy landscape $E(\theta)$:\cite{Tang2022NE,Yun2017SR,Timopheev2016}
\begin{equation}
\label{eq:PMAenergy}
E = -\frac{1}{2} \mu_{0} H_\mathrm{K1} M_\mathrm{S} \cos^{2}\theta
    -\frac{1}{4} \mu_{0} H_\mathrm{K2} M_\mathrm{S} \cos^{4}\theta
    -\mu_{0} H_{\parallel} M_\mathrm{S} \cos(\theta_\mathrm{H} - \theta)
\end{equation}
Here, $H_\mathrm{K1}$ and $H_\mathrm{K2}$ denote the first- and second-order PMA fields, respectively, while $\theta_\mathrm{H}$ represents a slight misalignment (less than 2$^{\circ}$, see Methods) of the applied $H_{\parallel}$ relative to the film plane. The equilibrium magnetization state is obtained at the minima of $E(\theta)$. The phase diagram in Figure \ref{fig:IPtransport}b shows the accessible magnetic states by tuning $H_\mathrm{K1}$ and $H_\mathrm{K2}$ (see Figure S10, Supporting Information). Figure \ref{fig:IPtransport}c presents the $\rho_{xy}$/$\rho_\mathrm{A}$ as a function of $H_{\parallel}$ measured at different temperatures for the 6 nm FGT/WSe$_2$ heterostructure. Due to $I \!/\!/\! H_{\parallel}$ (Figure \ref{fig:IPtransport}a), both the OHE and the planar Hall effect (PHE) are excluded, resulting in $\rho_{xy} \propto M_{\perp}$ and $\cos\theta$ = $\rho_{xy}$/$\rho_\mathrm{A}$. At low temperatures (2 and 200 K), a monotonic increase of $\rho_{xy}$/$\rho_\mathrm{A}$ with decreasing $H_{\parallel}$ from high field indicates a coherent rotation of the magnetization from IP to OP direction, further resulting in identical $\rho_{xy0}$ values for both OP and IP field sweeps. At elevated temperatures (250 and 300 K), the weaker PMA leads to a large number of multi-domain states forming near zero field, consequently reducing the net magnetization and $\rho_{xy}$. Furthermore, we extracted $\cos\theta(H_{\perp})$ = $\rho_{xy}(H_{\perp})$/$\rho_\mathrm{A}$ (blue curve) and $\sin\theta(H_{\parallel}) = \sqrt{1 - \left[\rho_{xy}(H_{\parallel})/\rho_\mathrm{A}\right]^2}$ (pink curve) at 2 K (Figure \ref{fig:IPtransport}c), corresponding to $M_{\perp}$ versus $H_{\perp}$ and $M_{\parallel}$ versus $H_{\parallel}$, respectively. The square and linear trends of these curves align with the SQUID data (Figure \ref{fig:SQUID}a). These results strongly confirm a robust, uniaxial PMA without metastable tilted states.

For quantitative analysis, $\rho_{xy}$/$\rho_\mathrm{A}$ was fitted (red lines, Figure \ref{fig:IPtransport}c) using the macrospin model of Equation (\ref{eq:PMAenergy}) (see Methods), valid in the single domain regime when the field decreases from a high field. Our analysis indicates that a higher-order PMA term ($H_\mathrm{K2}$) is essential to adequately capture the experimental response. Figure \ref{fig:IPtransport}d displays the extracted $H_\mathrm{K1}$ and $H_\mathrm{K2}$ as a function of temperature. Both decrease monotonically due to thermal fluctuations,\cite{Tang2022NE} yet remaining within the uniaxial PMA regime far from phase boundaries (symbols, Figure \ref{fig:IPtransport}b). These findings underscore the robustness of PMA over a wide temperature range. The emergence of higher-order PMA is likely due to the presence of multiple Fe sublattices in FGT, each experiencing distinct SOC and crystal field environments.\cite{Ershadrad2022JPCL,Bera2024PRB} This microscopic complexity aligns with the rich magnetic behaviors observed in FGT, including PMA-IMA transitions,\cite{Miao2023ACSnano} canted phases \cite{Tang2022NE,Zhao2023AM} and non-collinear spin order\cite{Schmitt2022CP}. In contrast, Fe$_3$GeTe$_2$ and Fe$_3$GaTe$_2$, which possess fewer Fe sites, exhibit a simpler magnetic behavior, exclusively characterized by PMA and devoid of temperature-induced phase transitions.\cite{Fei2018NM,Lopes2021,Shinwari2025} Meanwhile, the SOC in FGT is predominantly attributed to the Te atoms, while the Fe d-orbitals contribute moderate SOC strength.\cite{Ershadrad2022JPCL,Bera2024PRB} This aligns with the observation that Fe$_3$GeTe$_2$ and Fe$_3$GaTe$_2$, which possess a higher Te/Fe ratio and thus enhanced SOC, exhibit stronger PMA than FGT.\cite{Lopes2021,Kim2021ACSnano,Shinwari2025} Nevertheless, the more complex PMA landscape in FGT offers enhanced tunability of magnetic properties, offering a distinct advantage for tailoring magnetization dynamics, such as field-free SOT switching.\cite{Fert2024PRM,Zhao2025cis,Zhang2025cis} Finally, Figure \ref{fig:IPtransport}e compares the saturation fields $H_\mathrm{S}$ obtained from AHE measurements with IP (see estimation in Figure \ref{fig:IPtransport}c) and OP (Figure \ref{fig:AHE}a,b) fields. A significantly smaller $H_\mathrm{S}$ in the OP direction confirms the PMA up to room temperature, consistent with previous results from AHE (Figure \ref{fig:AHE}a-d) and SQUID (Figure \ref{fig:SQUID}) characterizations.

Figure \ref{fig:IPtransport}f shows the in-plane longitudinal magnetoresistance, defined as $MR_{\parallel}$ = [$\rho_{xx}(H_{\parallel})$ - $\rho_{xx}$(0 T)]/$\rho_{xx}$(0 T) × 100\%, with the indicated magnetization (pink arrows) and current (green arrows) directions. At 2 K, the positive $MR_{\parallel}$ observed in a large field ($M \parallel I$) compared to zero field ($M \perp I$) arises from the AMR effect. The extracted AMR of 5.8\% (definition in Figure \ref{fig:IPtransport}f) notably exceeds that of Fe$_3$GeTe$_2$ under similar conditions.\cite{You2019PRB} Given that AMR is typically correlated with the SOC strength,\cite{Ritzinger2023} this result is unexpected due to a weaker SOC in FGT than Fe$_3$GeTe$_2$. However, the AMR response in FGT may reflect the unequal contributions of its multiple Fe sublattices. Among them, Fe(1) experiences stronger SOC due to hybridization between the Fe 3d and Te 5p orbitals with a larger orbital magnetic moment.\cite{Ershadrad2022JPCL} The pronounced AMR observed in FGT can thus be primarily attributed to Fe(1), whose contribution dominates the spin-dependent scattering. This hypothesis is further supported by structural data: the Te–Fe(1) bond length in FGT (~2.61 $\mathrm{\AA}$) is shorter than that in Fe$_3$GeTe$_2$ (~2.73 $\mathrm{\AA}$),\cite{Hu2022} likely enhancing SOC at Fe(1) and contributing to the higher AMR in FGT. The temperature dependence of AMR further corroborates this perspective. Above 200 K, the AMR becomes negligible, despite only modest changes in both $M$ and $\rho_\mathrm{A}$ (Figure \ref{fig:SQUID}b and \ref{fig:AHE}c–d). This discrepancy may suggest that the significant suppression of AMR does not arise from reduced magnetization, but rather from magnetic unordering of Fe(1), leading to a diminished SOC-mediated scattering contribution. This assertion is further substantiated by the observation that $\rho_\mathrm{A}$ exhibits a decline around 110 K (Figure \ref{fig:AHE}d), a temperature much lower than the 0.75 $T_\mathrm{C}$ threshold expected for a magnetization-driven decrease.\cite{Nagaosa2010,Lv2023AS} Collectively, these observations suggest that carrier transport in FGT is predominantly governed by spin-dependent scattering at Fe(1), which exhibits substantial SOC, while the contributions from Fe(2) and Fe(3) remain limited due to their weaker SOC.

\subsection{\label{sec:leve26} Thickness-dependent unconventional Hall effect}
\begin{figure*}[t!]
\centering
\includegraphics*[width=0.8\textwidth]{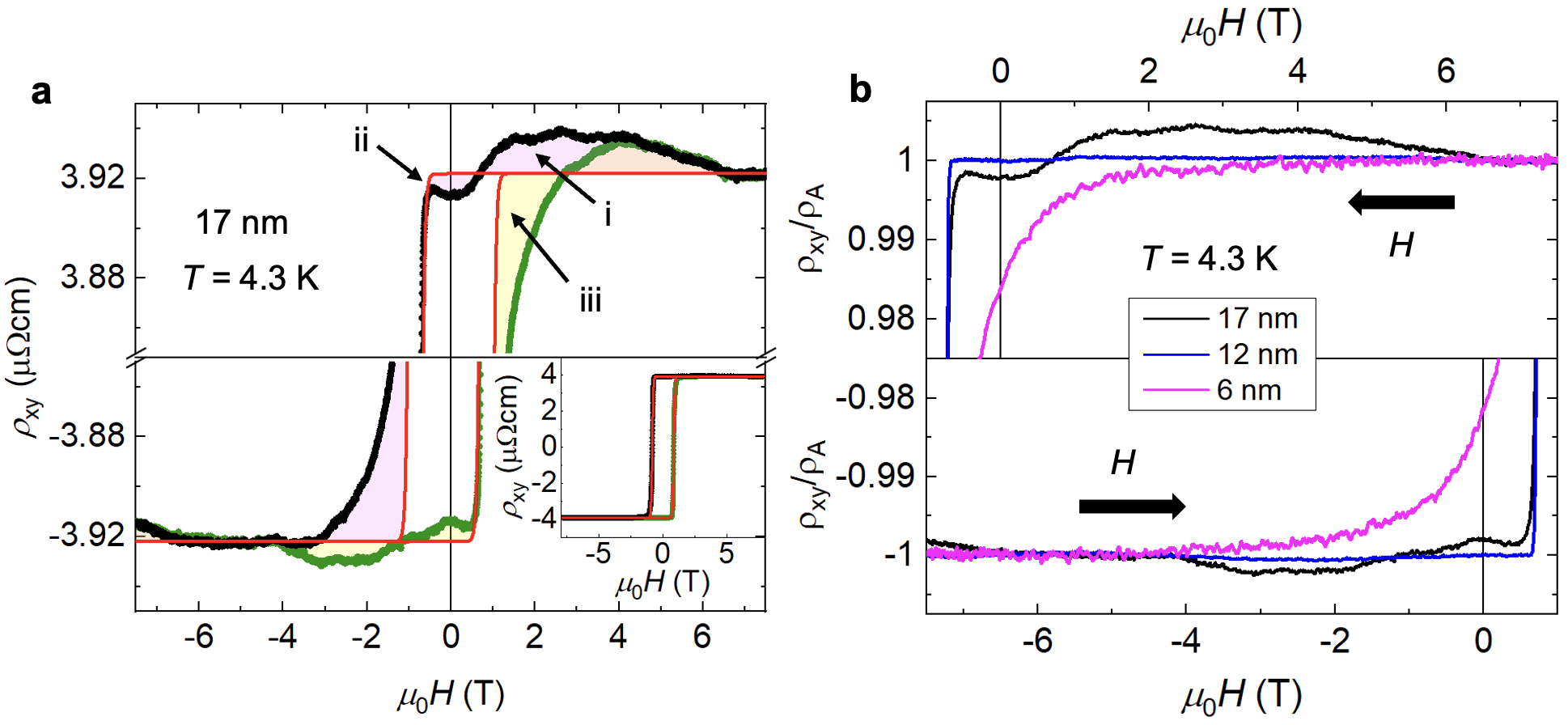}
\caption{Unconventional Hall effect (UHE) in FGT/WSe$_2$ vdW heterostructures.
a) $\rho_{xy} - H$ of a 17 nm FGT/WSe$_2$ sample at 4.3 K with downward (black) and upward (green) field sweeps after subtraction of the linear OHE contribution. The red curves represent the assumed conventional AHE component, and the shaded regions highlight the unconventional contributions. Inset: full $\rho_{xy} - H$ loop showing the sharp magnetization switching.
b) Thickness dependence of the UHE. For clarity, only the curves obtained while reducing the magnetic field from high values are shown, with sweep direction indicated by black arrows.} 
\label{fig:THE}
\end{figure*}

To evaluate the potential formation of topological magnetic orders in our FGT/WSe$_2$ heterostructures, the AHE measurements were further performed using a high-resolution setup, as shown in \textbf{Figure \ref{fig:THE}}a for a 17 nm FGT/WSe$_2$ sample measured at 4.3 K. 
Indeed, a clear unconventional Hall effect (UHE) was observed during both forward and backward field sweeps (highlighted shaded regions), deviating from the conventional AHE baseline (red curves). Additional UHE results for different Hall devices and temperatures can be found in our previous report.\cite{Lv2024} Three distinct deviations with characteristic up-down-up features are consistent with topological Hall effect (THE) signatures arising from topological spin textures\cite{Lv2024SkrAFM,Sivakumar2020,Zhang2022ASskyr,Wu2020NC}, well matching theoretical predictions for the influences of skyrmions and antiskyrmions on Hall signals.\cite{Sivakumar2020} In fact, the skyrmion formation and corresponding THE are highly expected in FGT/WSe$_2$ heterostrucures due to their intrinsic magnetic properties and the potential influence of interfacial proximity effects.\cite{Schmitt2022CP,Lv2024SkrAFM,Zhang2022ASskyr,Wu2020NC} 
Importantly, the possibility that the observed AHE anomalies arise from additional magnetic phases can be confidently excluded. For this hypothesis to be valid, the hypothetical secondary phase must produce an AHE with a sign opposite to FGT, thereby requiring distinct Berry curvatures. However, such a phase would manifest as distinct magnetic behavior, including different coercive fields and multi-step, narrow-waist magnetization switching. In contrast, the full-range loops exhibit sharp single-step switching (inset in Figure \ref{fig:THE}a), without indication of multiple discrete transitions. This conclusion is further supported by in-plane field measurements (Figure \ref{fig:IPtransport}c), which demonstrates a continuous and seamless rotation of the magnetization from IP to OP directions. These findings may suggest the formation of skyrmions within our FGT/WSe$_2$ heterostructures, with the observed UHE attributed to THE. We anticipate that further enhancement and clearer observation of THE will be achievable in nanoscale Hall bar devices with lateral dimensions comparable to skyrmion sizes. Combined with skyrmion imaging, this approach could enable detailed studies of skyrmion–carrier interactions, which we plan to pursue systematically in future work.

To investigate the influence of WSe$_2$ on UHE, we systematically examined its dependence on FGT thickness. For clarity, Figure \ref{fig:THE}b only presents the curves recorded during the magnetic field sweep from high to low field, with black arrows indicating the sweep direction. In contrast to the 17 nm FGT sample (black), which exhibits a clear UHE signature, the 12 nm (blue) and 6 nm (pink) samples manifest only conventional AHE. Notably, the detection of UHE in transport measurements critically depends on the skyrmion density. A threshold density is required to produce a measurable UHE signal.\cite{Lv2024SkrAFM} According to a recent study on skyrmions in FGT, a topological Hall resistivity of 0.02 $\mathrm{\mu} \ohm$cm [region (i) in Figure \ref{fig:THE}a] corresponds to a skyrmion density about 1 $\mathrm{\mu}$m$^{-2}$.\cite{Lv2024SkrAFM} Assuming a skyrmion size of 65 nm,\cite{Lv2024SkrAFM} this yields a ratio of 0.004 between the skyrmion area and the total FGT area. Notably, this value closely matches the ratio of the UHE and AHE signals ($\approx$ 0.004, Figure \ref{fig:THE}b). This consistency further supports the correlation between the UHE and skyrmion formation in our samples. Consequently, the absence of the UHE signal in thinner samples may not preclude skyrmion formation but likely indicates a lower skyrmion density compared to thicker films. These findings imply that skyrmions predominantly form within the FGT bulk rather than at the FGT/WSe$_2$ interface. This contrasts with Fe$_3$GeTe$_2$-based heterostructures, such as Fe$_3$GeTe$_2$/WTe$_2$, where skyrmions were reported to form at the heterointerface rather than within the bulk.\cite{Wu2020NC} A potential origin of this difference may lie in the different PMA strengths of Fe$_3$GeTe$_2$ and FGT. Typically, Fe$_3$GeTe$_2$ exhibits a stronger PMA,\cite{Fei2018NM,Lopes2021} which favors collinear spin alignment and thus suppresses skyrmion formation. In such systems, the adjacent TMDC layer can modulate PMA to an intermediate regime that promotes non-collinear spin textures at the interface. In contrast, FGT films inherently exhibit moderate PMA strength, allowing skyrmion formation.\cite{Lv2024SkrAFM,Schmitt2022CP} However, the presence of WSe$_2$ may further enhance PMA due to its large SOC and interfacial proximity effects, potentially impeding skyrmion nucleation in thinner FGT. Therefore, PMA acts as a key parameter for controlling the balance between collinear and non-collinear spin configurations, critically influencing skyrmion formation. This provides FGT with a distinct advantage over Fe$_3$GeTe$_2$ on skyrmion creation. Moreover, while interfacial skyrmions are typically 2D spin textures, skyrmions in thicker FGT layers could support more complex topological spin texture, such as 3D hopfion rings.\cite{Zhou2025AM,Zheng2023N} This expands the landscape of accessible magnetic textures. The formation and density of skyrmions in FGT/WSe$_2$ heterostructures may be further manipulated by tuning temperature, magnetic field and layer thickness. In addition, skyrmions in FGT can be topologically protected, providing robust stability against external perturbations.\cite{Casas2023AM,Schmitt2022CP} These potentials make FGT/WSe$_2$ heterostructures a compelling platform for skyrmion research and applications, including the skyrmion Hall effect,\cite{Litzius2017} skyrmion-based racetrack memory\cite{Fert2013,Parkin2015} and neuromorphic computing\cite{deCamara2025}.

\begin{figure*}[t!]
\centering
\includegraphics*[width=0.45\textwidth]{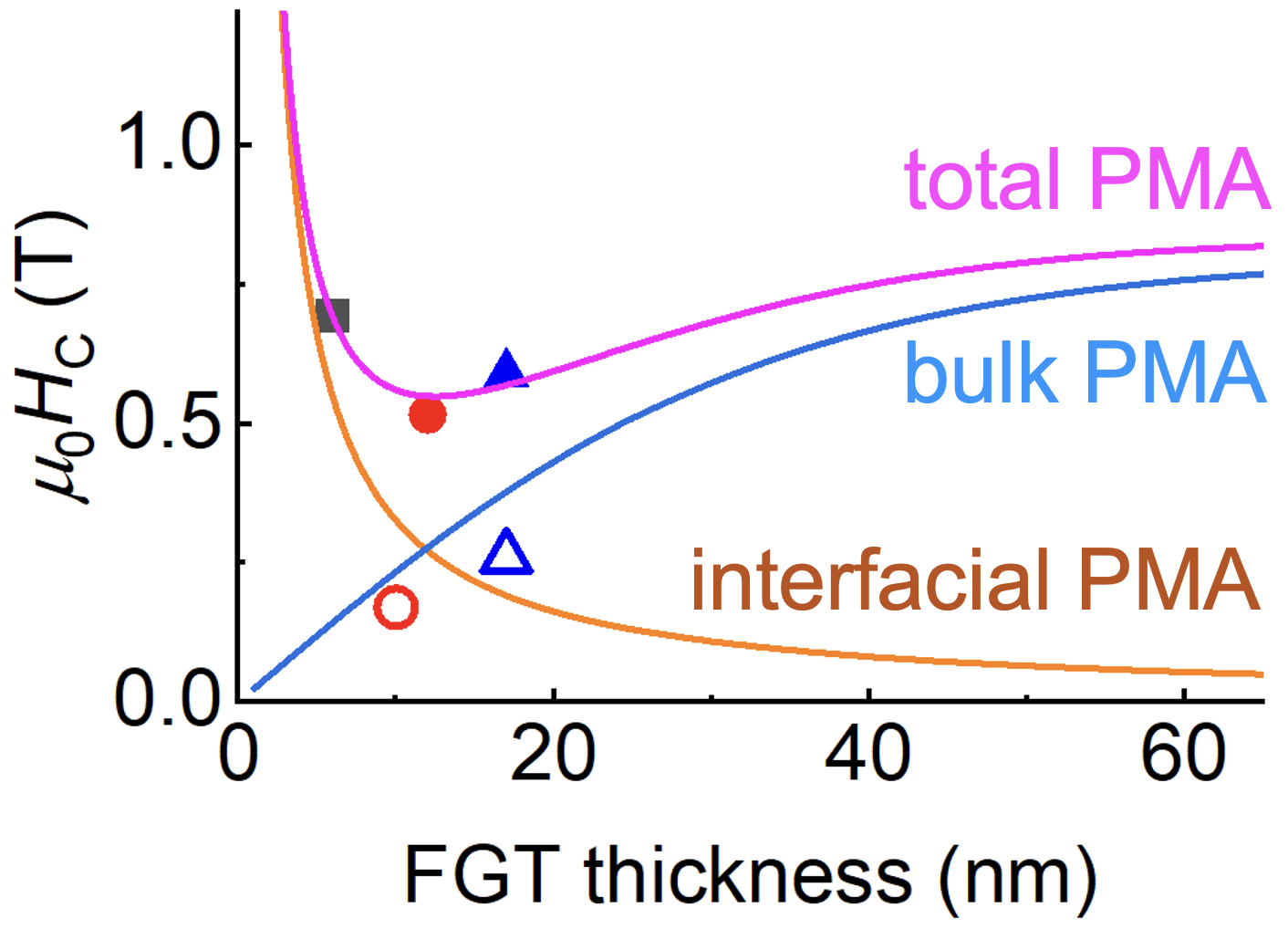}
\caption{Non-monotonic thickness dependence of $H_\mathrm{C}$ in FGT/WSe$_2$ heterostructures (filled symbols). The data are fitted (pink curve) using Equation (\ref{eq:hc}), with contributions from bulk PMA (blue curve) and interfacial PMA (orange curve). Open symbols show reference data obtained from FGT/graphene heterostructures.\cite{Lv2023Small}
}
\label{fig:hc}
\end{figure*}

\subsection{\label{sec:leve27} Proximity-induced interfacial contributions to magnetic and transport anomalies}
To further investigate the influence of WSe$_2$ on the magnetic and transport properties of FGT, we examined the thickness dependence of $H_{\mathrm{C}}$ for FGT films (taken from AHE curves in Figure \ref{fig:AHE}a at 4.3 K) grown on epitaxial WSe$_2$ and, as a reference, on epitaxial graphene.\cite{Lv2023Small,Lv2024,Lopes2024} Unlike the monotonic behavior typically reported for exfoliated FGT flakes,\cite{Alghamdi2024,Kim2021ACSnano} $H_{\mathrm{C}}$ in FGT/WSe$_2$ exhibits an unconventional non-monotonic dependence on $t_{\mathrm{FGT}}$ (\textbf{Figure \ref{fig:hc}}, filled symbols). This behavior can be well described by considering both bulk and interfacial PMA contributions:\cite{Kim2021ACSnano}
\begin{equation}
\label{eq:hc}
H_{\mathrm{C}} = H_{\mathrm{C0}} \, \tanh\left(\frac{2 \, t_{\mathrm{FGT}}}{t_0}\right) + \frac{J_{\mathrm{int}}}{M_{\mathrm{S}} \, t_{\mathrm{FGT}}}
\end{equation}
The first term, saturating at $H_{\mathrm{C0}}$ for $t_{\mathrm{FGT}}>t_0$, describes bulk PMA with a feature typically reported in 2D ferromagnet flakes.\cite{Alghamdi2024,Kim2021ACSnano} The second term, proportional to $1/t_{\mathrm{FGT}}$, captures interfacial PMA from interlayer coupling at the FGT/WSe$_2$ interface. Fitting yields $t_0$ = 66 nm, consistent with values reported for Fe$_3$GeTe$_2$ ($\approx$ 60 nm),\cite{Kim2021ACSnano} and $J_{\mathrm{int}} \approx$ 1.0 mJ/m$^2$, comparable to conventional interfacial PMA systems such as Co/Pt (0.6 $\sim$ 1.4 mJ/m$^2$)\cite{Guo2006,Yakushiji2010,Dieny2017RMP} and CoFeB/MgO (1.8 mJ/m$^2$).\cite{Worledge2011,Dieny2017RMP}
The pronounced $1/t_{\mathrm{FGT}}$ term reveals a strong interface-driven enhancement of PMA. In 3D ferromagnet/heavy-metal and ferromagnet/oxide systems, the interfacial PMA typically originates from SOC–mediated hybridization at the interface.\cite{Ikeda2010,Ikeda2012,Dieny2017RMP} By analogy, our results strongly indicate that the interfacial PMA in FGT/WSe$_2$ is attributed to SOC from the WSe$_2$ underlayer — a signature of a proximity effect. This interpretation is further supported by the consistently higher $H_{\mathrm{C}}$ in FGT/WSe$_2$ (filled symbols, Figure \ref{fig:hc}) compared to FGT/graphene (open symbols, Figure \ref{fig:hc}), despite similar crystalline quality,\cite{Lv2023Small,Lv2024,Lopes2024} in line with the much stronger SOC in WSe$_2$ than in graphene. Furthermore, our results align well with previous reports of proximity-induced enhancement of $H_\mathrm{C}$ through TMDC integration.\cite{Ma2024}
The non-monotonic PMA profile, arising from the competition between bulk and interfacial contributions, is a distinctive feature of vdW heterostructures. In contrast, exfoliated FGT flakes exhibit only the bulk term,\cite{Alghamdi2024,Kim2021ACSnano} while 3D ferromagnet/heavy-metal and ferromagnet/oxide systems show only the interfacial contribution.\cite{Dieny2017RMP} This integration provides enhanced tunability of magnetic anisotropy, which could be further modulated via the WSe$_2$ thickness.
Overall, the thickness scaling, quantitative fits, comparison between FGT/WSe$_2$ and FGT/graphene, as well as SOC considerations provide strong evidence that the magnetic and transport anomalies (e.g, unusual thickness dependence of UHE and PMA) in our FGT/WSe$_2$ heterostructures arise from an interfacial SOC-driven proximity effect, enabled by the clean and atomically sharp nature of the interface. This tunability establishes FGT/WSe$_2$ as a promising platform for studying proximity-driven physics.

\section{Conclusion}
In this work, we have demonstrated the realization of large-scale, all-epitaxial FGT/WSe$_2$ vdW heterostructures exhibiting unconventional magnetic and transport properties unique to their 2D nature. 
Large-area synthesis using MBE and CVD enables us to overcome the scalability challenge, achieving ferromagnetic/semiconducting vdW heterostructures that exhibit room temperature ferromagnetism. This marks a significant advance in the realization of 2D magnetic systems for device integration. 
The FGT/WSe$_2$ heterostructures exhibit PMA
while maintaining the $T_\mathrm{C}$, which is beneficial  for low-power spintronic applications. The unusual thickness- and temperature-dependent sign reversal of exchange bias highlights distinctive 2D magnetic behavior. In-plane magnetotransport further reveals a pronounced higher-order PMA contribution and unusually large AMR effect, pointing to sublattice-specific roles in governing magnetism and transport unique to FGT. The reversed thickness dependence of the UHE, opposite to trends reported for Fe$_3$GeTe$_2$-based heterostructures, underscores the interplay between PMA and interfacial coupling, suggesting the possible formation of skyrmions. Together with the non-monotonic PMA-thickness dependence, these magnetic and transport anomalies can be attributed to a strong interfacial proximity effect at the FGT/WSe$_2$ interface, enabled by the high epitaxial quality and the strong SOC of WSe$_2$. These heterostructures establish a robust and versatile platform for both fundamental studies and next-generation 2D spintronic technologies, including vdW magnetoresistive devices, ultrahigh-density racetrack memories and proximity-engineered magnetic functionalities.

\section*{Experimental Section}  

\threesubsection{MBE growth} FGT was grown by MBE using elemental Fe, Ge, and Te evaporated from Knudsen cells. The flux for each element was obtained by measuring the beam equivalent pressure employing a pressure gauge. For the flux ratios utilized in this study, the average composition of the films is Fe$_{4.8}$GeTe$_2$.\cite{Lv2023Small} Continuous films with thicknesses of around 6, 12, and 17 nm were prepared at a substrate temperature of 260 $^\circ$C. In-situ growth monitoring was performed by RHEED. As growth templates, we used single-crystalline, continuous monolayer–bilayer thick WSe$_2$ films synthesized on Al$_2$O$_3$(0001) by CVD. Details about the CVD growth of WSe$_2$ can be found elsewhere.\cite{Zhu2023NatNano} Prior to FGT growth, the 1 cm$^2$ large WSe$_2$/Al$_2$O$_3$(0001) templates (cut out of 2-inch wafers) were in situ annealed at 300 $^\circ$C for 20 minutes in order to remove surface contaminants. The FGT films were capped in-situ after their growth with a Te layer (5 nm) deposited after sample cooling to room temperature. This procedure was adopted in order to minimize FGT surface oxidation upon air exposure. This provided additional protection against FGT oxidation during transportation of the samples from the MBE laboratory in Germany to the STEM laboratory in Spain.

\threesubsection{Grazing incidence diffraction} Measurements shown in Figure \ref{Fig:GID} were performed at beamline P23 at PETRA III using an X-ray energy of 18 keV. Experiments shown in Figure S3 (Supporting Information) were performed at the BM25- SpLine beamline at The European Synchrotron (ESRF) in Grenoble. An incidence angle of the illuminating X-rays of 0.2$^\circ$ sufficiently suppresses strong scattering by the substrate, and thus makes this method highly surface sensitive. An X-ray wavelength of 0.689 \AA{} (corresponding to a photon energy of 18 keV), selected by a Si(111) monochromator, enables inspection of a comparatively large area in reciprocal space, which is important for accessing multiple reflections of the same lattice plane family.

\threesubsection{Scanning transmission electron microscopy (STEM) and electron energy loss spectroscopy (EELS)} An electron transparent TEM lamella was prepared in a Dual Beam Helios 5 UX Focused ion beam.
Atomic-resolution high-angle annular dark-field (HAADF) STEM images were acquired in a double-corrected Thermofisher SPECTRA 300 (S)TEM microscope operated at 200KV. For the image acquisitions, a convergence and collection semiangles of around 20 mrad and 67-200 mrad were used. EELS spectrum images were acquired using a continuum spectrometer from GATAN, equipped with the K3 direct electron detection camera and operated in counted mode, 40-50 mrad collection angle, 20 pm pixel size, 70 pA probe current and 2 ms dwell time. 

\threesubsection{Magnetic characterization} The magnetic properties of the FGT/WSe$_2$ films were characterized using a SQUID magnetometer (Quantum Design MPMS-3) under vacuum with magnetic fields up to 5 T. Hysteresis loops (Figure \ref{fig:SQUID}a) were acquired by sweeping the external magnetic field, and a linear diamagnetic background was subtracted (Figure S5, Supporting Information). For the $M$–$T$ measurements (Figure \ref{fig:SQUID}b), the samples were first cooled to 4 K in zero field, after which the magnetization was initialized using a 1 T field. The field was then reduced to zero, and the $M$–$T$ curves were recorded during subsequent zero-field warming. All measurements were performed for both out-of-plane (OP) and in-plane (IP) field configurations.

\threesubsection{Hall device fabrication} Thin FGT/WSe$_2$ films were patterned into Hall bar devices using standard cleanroom fabrication techniques. Since Ar$^{+}$ plasma etching makes the photoresist difficult to remove, the electrodes were first defined by optical lithography, followed by electron beam evaporation of Cr (9 nm)/Au (80 nm). The Hall bar structures were then defined through a second optical lithography step and Ar$^{+}$ ion-beam etching. To prevent peeling of vdW layers during photoresist removal, a hot acetone bath at 50 $^{\circ}$C was employed instead of the conventional lift-off process. Finally, the fabricated Hall devices were wire-bonded to a chip carrier using Al wires. Figure  S7 (Supporting Information) displays an image of fabricated Hall device, with a width of 100 $\mathrm{\mu}$m and a length between two arms of 300 $\mathrm{\mu}$m.

\threesubsection{Magnetotransport} AHE measurements (Figure \ref{fig:AHE}) were performed on Hall bar devices (Figure S7, Supporting Information) under external out-of-plane magnetic fields up to 0.8 T in vacuum conditions ($10^{-6}$ to $10^{-7}$~mbar). The AHE loops were initially measured at room temperature, followed by zero-field cooling to 4.3 K. Subsequent AHE curves were measured at each temperature from 4.3 to 400 K. A constant direct current (DC) was applied using a Hewlett-Packard 3245A current source, while Hall voltage was measured using a Hewlett-Packard 3458A digital multimeter. In-plane transport measurements (Figure \ref{fig:IPtransport}) were performed with an in-plane field up to 7 T parallel to the current direction (see Figure \ref{fig:IPtransport}a) from 2 to 300 K. UHE measurements (Figure \ref{fig:THE}) were performed in an Attocube system with an out-of-plane field up to 8 T and a lock-in detection (SR 830) at a frequency of $\approx$ 37 Hz.

\threesubsection{Data simulation} For the analysis of the in-plane transport data (Figure \ref{fig:IPtransport}c), the stable magnetization state was obtained from Equation (\ref{eq:PMAenergy}) under the conditions of $\partial E/\partial \theta = 0$ and $\partial^2 E/\partial \theta^2 > 0$, resulting in:
\begin{equation}
\label{eq:e0}
H_{K1} \cos\theta \sin\theta + H_{K2} \cos^3\theta \sin\theta = H_\parallel \sin(\theta_H - \theta)
\end{equation}
The transport data were fitted using Mathematica, as shown by red curves in Figure \ref{fig:IPtransport}c. The best fit results in $\theta_H \approx 88^\circ$, indicating a slight $\approx 2^\circ$ deviation of the applied field from the film plane. See also Figures S11, Supporting information.
For the analysis of thickness-dependent $H_{\mathrm{C}}$ using Equation (\ref{eq:hc}) (Figure \ref{fig:hc}), we used $M_{\mathrm{S}}$ = 308 emu/cm$^{3}$ estimated from SQUID measurements in the saturation regime (4–5 T in Figure \ref{fig:SQUID}a). Fitting the data yields $J_{\mathrm{int}} \approx$ 1.0 mJ/m$^2$, $t_0$ = 66 nm and $H_{\mathrm{C0}}$ = 0.8 T.

\medskip
\textbf{Supporting Information} \par 
Supporting Information is available from the Wiley Online Library or from the author.

\medskip
\textbf{Acknowledgements} \par 
We thank Dr. A.Khadiev and Dr. D. Novikov for their support during setting up the experiment. We also acknowledge PETRA III for providing beamtime through proposal I-20230728. We thank the ESRF for providing beamtime through proposal HC-6285 and Dr. Juan Rubio Zuazo for his assistance during the experimental setup. 
Authors acknowledge the use of instrumentation as well as the technical advice provided by the Joint Electron Microscopy Center at ALBA (JEMCA) and funding from Grant IU16-014206 (METCAM-FIB) to ICN2 funded by the European Union through the European Regional Development Fund (ERDF), with the support of the Ministry of Research and Universities, Generalitat de Catalunya. Authors acknowledge Marcos Rosado for the lamella preparation.
The group in Mainz acknowledges the German Research Foundation (DFG SFB TRR173 Spin+X 268565370, projects A01, B02 and A12 and DFG SFB TRR288 ELASTO-Q-MAT 422213477, projects A09 and A12),  the European Union’s Horizon 2020 Research and Innovation Programme under grant agreement 856538 (project “3D MAGIC”) and King Abdullah University of Science and Technology (KAUST) under award 2024-CRG12-6480.

\medskip

\bibliographystyle{MSPtitle}
\bibliography{FGTWSe2}


\end{document}